\documentclass[preprint,aps,floats,nofootinbib]{revtex4}

\newcommand{\be}{\begin{equation}}
\newcommand{\ee}{\end{equation}}
\newcommand{\bea}{\begin{eqnarray}}
\newcommand{\eea}{\end{eqnarray}}

\def\thebibliography#1{\centerline{\bf REFERENCES}
  \list{[\arabic{enumi}]}{\settowidth\labelwidth{[#1]}\leftmargin
  \labelwidth\advance\leftmargin\labelsep\usecounter{enumi}}
\def\newblock{\hskip .11em plus .33em minus -.07em}\sloppy
  \clubpenalty4000\widowpenalty4000\sfcode`\.=1000\relax}

\begin{document}
\draft

\vspace*{0.5cm}

\title{Generalized Study with Isospin-phased Topological Approach \\
on the $B \to K\pi$ Puzzle\footnote{Talk given at the third International
Conference on Flavor Physics(ICFP 2005), National Central University,
Chong-li, Taiwan, R.O.C, 3-8 Oct 2005. (Web: {\tt http://www.phy.ncu.edu.tw/hep/icfp2005}).}}

\author{ \vspace{0.5cm}
D.~Chang$^{1,4}$,
C.~S.~Chen$^{1,4}$\footnote{E-mail: d927305@oz.nthu.edu.tw}
H.~Hatanaka$^{1,4}$\footnote{E-mail: hatanaka@phys.nthu.edu.tw},
C.~S.~Kim$^{2,3}$\footnote{E-mail: cskim@yonsei.ac.kr} ~
}

\affiliation{ \vspace{0.3cm}
$^{1}$ Department of Physics, National Tsing-Hua University,
Hsinchu 30013, Taiwan, R.O.C.
\\
$^2$ Department of Physics, Yonsei University,
Seoul 120-749, Korea
\\
$^3$ Graduate School of Sciences, Hiroshima University,
Higashi-Hiroshima, Japan
\\
$^4$ Physics Division, National Center for Theoretical Sciences, Hsinchu
30013, Taiwan, R.O.C
\vspace{1cm}}

\vspace*{0.5cm}

\begin{abstract}
\noindent We study the decay processes $B \to K \pi$ by using
generalized decay amplitudes with both the final state
re-scattering strong phases and the topological quark
diagrammatic strong phases included together as part of fitting
parameters.
Using a generalized approach, so called ``isospin phased topological
approach'', for all the currently available data of $B \to K \pi$ decays,
we determine the
allowed values of the relevant theoretical parameters, corresponding to
the electroweak penguin, the color-suppressed tree contribution,
strong phase differences, etc.
In order to find the most likely values of the parameters in a
statistically reliable way, we use the $\chi^2$ minimization technique.
We find that the long distance final state re-scattering, when taken at
proper value, can provide a reasonable fit to the standard model with the
perturbative QCD estimated values,
and therefore, it is
premature to conclude that it requires new physics to explain the CP
violating $B \to K\pi$ data.
%
\end{abstract}

\maketitle

\section{Introduction}

There are four different decay channels (and their anti-particle decay
channels) for $B \to K\pi$ processes, depending on the electric charge
configuration: $B^+ \to K^0 \pi^+$, $B^+ \to K^+ \pi^0$, $B^0 \to K^+
\pi^-$, and $B^0 \to K^0 \pi^0$.
All the $B \to K\pi$ modes have already been observed in experiments and
their CP-averaged branching ratios have been measured within a few
percent errors by the BaBar and Belle collaborations
\cite{Bornheim:2003bv,Chao:2003ue,Aubert:2002jb,Aubert:2004dn,Aubert:2004km,
Aubert:2004kn}.
The observations of the direct CP asymmetry in $B^0 \to K^{\pm}
\pi^{\mp}$ have also been recently achieved at the 5.7$\sigma$ level by BaBar
and Belle \cite{Chen:2000hv,Aubert:2004qm,Chao:2004jy,Chao:2004mn,Abe:2004xp}.
For the other $B \to K\pi$ modes, the experimental results of the direct
CP asymmetries still include large errors.  Certain experimental data
($e.g.$, the branching ratios (BRs)) for $B \to K\pi$ are currently more
precise than the theoretical model predictions based on QCD
factorization (QCDF), perturbative QCD (pQCD), and so on.
Thus, these decay modes can provide very useful information for
improving the model calculations, and   at the same time, the
model-independent study becomes very important.


In the light of those new data, including the direct CP asymmetry in
$B^0 \to K^{\pm} \pi^{\mp}$, many works have been done to study
the implications of the data
\cite{Mishima:2004um,Buras:2004th,Charng:2004ed,He:2004ck,Wu:2004xx,
Baek:2004rp,Carruthers:2004gj,Nandi:2004dx,Morozumi:2004ea,Kim:2005jp}.
The quark level subprocesses for $B \to K\pi$ decays are $b \to s q \bar
q ~ (q = u,d)$ penguin processes, which are potentially very sensitive to any
new physics effects beyond the standard model (SM).  Thus, with the
currently available precision data, it is very important to investigate
these modes as generally and critically as possible.

At the bottom of this important problem, lies the fact that final state
re-scattering phases by strong interaction play a crucial role in
generating CP violation.
Part of the strong interaction phases (as short-distance strong phases
(sdSPs) related to topological quark diagrams) can be investigated by using various
QCD models such as QCDF or pQCD,
even though the results are quite model dependent.
However, there are also long-distance final state strong phases (ldSPs),
that are very difficult to calculate due to
hadronic interactions at low energy scale
(even thought some attempts had been done).
Despite facing  the fact that we do not
understand strong phases very well and that there are already many
theoretical claims of possible new physics from the data of $B \to K \pi$,
here we are trying to
be as model-independent as
possible in fitting the $B \to K\pi$ data, by choosing both sdSPs
and ldSPs  as free fitting parameters,
to see if new physics is still required by the recent data or not.

In this work, we study the decay processes $B \to K\pi$ by using
generalized decay amplitudes within the so-called ``isospin-phased
topological approach'', $i.e.$ with both the strong phases
of the final state re-scattering (ldSPs) and the strong phases of
the topological quark diagrammatic origin (sdSPs).
If we ignore topological strong phase
differences, this general approach becomes  the original isospin approach.
Inversely, if
we ignore the strong phase differences in isospin amplitudes, it reduces
to the normal topological analysis \cite{Gronau:1994rj}, which is based
on an exact flavor SU(3) symmetry in $B$ meson decays
\cite{Zeppenfeld:1980ex}.

We note that the importance of final state interactions (FSI) has
recently been recognized again in hadronic $B$ decays
\cite{Cheng:2004ru}, in which the importance of the
existence of the soft final state re-scattering effects has been pointed out,
especially in
a model-independent topological quark diagrammatic analysis of $B \to D
\pi$ \cite{Kim:2004hx} and $B \to K \pi$ decays.
It is certainly conceivable that FSI can modify the predictions based on
the short distance diagrammatic analysis.  Likewise, the branching ratios
of certain decay modes $B_{s,d} \to \pi \pi$ are expected to be
extremely small if re-scattering effects through FSI do not alter the
predictions of the diagrammatic approach.  We notice that the soft final
state re-scattering effects can violate the exact flavor SU(3) symmetry
even though the isospin symmetry will still hold with FSI, since FSI may
rather be at low energy scale of light final particles.
In other words, at the scale that FSI are activated, the flavor SU(3) is broken
but the SU(2) isospin symmetry is still valid. And FSI can be parameterized as the
isospin phases in the limit of the elastic re-scattering.

We note that including both sdSPs and ldSPs may amounts to a possible double
counting:
Since they typically involve physics
at different scales with different symmetries like
flavor SU(3) symmetry or SU(2) isospin symmetry,
the double counting may not be as serious as one thinks.
As is well known, the scattering of hadrons exhibits a two-component structure
of ``soft" and ``hard". We associate the high scale hard scattering components between
pointlike constituents with the SU(3) strong phases of quark diagramatic origin,
and the low scale soft components of FSI with the SU(2) isospin phases.
In addition, since they are just taken as model independent fitting
parameters, even if there are some double counting of physical effects
contained in our fitting parameters, it is not going to affect our final
physical conclusions regarding new physics.

Here we are mainly interested in investigating whether the conventional
SM predictions are consistent with the current data even after we
include FSI effects.  Furthermore, if there are some deviations between
the conventional estimates and the experimental results, we intend to
identify carefully the source of the deviations and estimate how large
the contribution from the source can be.
Then, by comparing our result with the conventional SM predictions, we
shall be able to verify whether the current data indicate any new
physics effects.  In order to find the most likely values of the
theoretical parameters in a statistically reliable way, we will adopt the
$\chi^2$ analysis.

We organize the paper as follows: In Section II, we introduce all the
relevant formulas for $B \to K \pi$ decays, step-by-step generalizing
the decay amplitudes.  We also give formulas for BRs, direct CP
violations and mixing-induced direct CP violation. In Section III, we
present the summary of the recent experimental results on $B \to K \pi$
modes. And we do $\chi^2$ analysis for all the experimental observables,
and discuss its physical implication within our generalized approach. In
Section IV, we give conclusions.

\section{The detailed formulas for $B \to K \pi$ decay modes in the isospin-phased
topological approach}
\renewcommand{\Re}{\mathop{\mathrm{Re}}\nolimits}
\renewcommand{\Im}{\mathop{\mathrm{Im}}\nolimits}
\renewcommand{\arg}{\mathop{\mathrm{arg}}\nolimits}
\newcommand{\lambdat}{{\lambda_{t}}}
\newcommand{\lambdau}{{\lambda_{u}}}
\newcommand{\strongT}{{e^{i\delta_T}}}
\newcommand{\strongPEW}{{e^{i\delta_{EW}}}}
\newcommand{\strongC}{{e^{i\delta_C}}}
\newcommand{\strongPCEW}{{e^{i\delta^C_{EW}}}}
\newcommand{\strongP}{{e^{i\delta_P'}}}
\newcommand{\deltaT}{{\delta_T}}
\newcommand{\deltaEW}{{\delta_{EW}}}
\newcommand{\deltaC}{{\delta_{C}}}
\newcommand{\alphaone}{{\alpha_{1/2}}}
\newcommand{\alphathree}{{\alpha_{3/2}}}
\newcommand{\pcew}{{P^C_{EW}}}
\newcommand{\pew}{{P_{EW}}}
\newcommand{\tfrac}[2]{{\textstyle \frac{#1}{#2}}}
We first introduce the decay amplitudes within the generalized
``isospin-phased topological approach", then we summarize the formulas
for the relevant decay amplitudes, BRs, direct and indirect
(mixing-induced) CP asymmetries of $B \to K \pi$ processes.

\subsection{Fusion of Isospin and Topological Approaches}

We write $B$-meson decay amplitude, $e.g.$ $B^0 \to \pi^- K^+$,
by including both the topological strong phases
and the isospin related strong phases,
so called ``isospin-phased topological approach",
as:
\begin{eqnarray}
{\cal A} &=& \sum_I A_I \exp(i \Delta_I) \\
&=& \sum_I[D_1 \exp(i \delta_1) + D_2 \exp(i \delta_2) + ...]_I \exp(i \Delta_I),
\label{eq-2}
\end{eqnarray}
where $A_I$ is an isospin amplitude with the isospin related (ldSPs)
strong phase $\Delta_I$,
and we decompose each isospin amplitude into
a sum of the topological diagrammatic  amplitudes, $D_i$,
with the topological strong phases (sdSPs) $\delta_i$.
As can be seen below, for $B^0 \to \pi^- K^+$ decay channel
$A_I$'s are expressed with 3 isospin amplitudes $A_{3/2},~A_{1/2}$ and $B_{1/2}$,
with  strong phases $\alpha'_{3/2},~\alpha'_{1/2}$ and $\beta'_{1/2}$,
respectively. And
each $A_I$ is expressed with a sum of the topological amplitudes, $T,C,P,P_{EW}$ etc.
with strong phases $\delta'_T,\delta'_C,\delta'_P,\delta'_{EW}$ $etc$., respectively.
Now if we ignore phase differences in the topological amplitudes,
we recover  the original
isospin amplitude with a positive real value $|A_I|$
(except for the relevant weak phases).
Inversely, if we ignore phase differences among isospin
amplitudes, the framework reduces to the normal topological analysis.

As final states, we take
$\pi^+ \equiv u\bar{d},
\pi^0 \equiv \frac{d\bar{d} - u\bar{u}}{\sqrt{2}},
\pi^- \equiv -d\bar{u},
$
for pions and
$
K^+ \equiv u\bar{s},
K^0 \equiv d\bar{s},
\bar{K}^0 \equiv s\bar{d},
K^- \equiv -s\bar{u},
$ for kaons, and for $B$ mesons
$
B^+ \equiv \bar{b}u,
B^0 \equiv \bar{b}d,
\bar{B}^0 \equiv b\bar{d},
B^- \equiv -b\bar{u}
$.
Isospin relations are
\begin{eqnarray}
A(B^0 \rightarrow \pi^- K^+)
& \equiv X =& A_{3/2} + A_{1/2} - B_{1/2},
\label{isospin-formulas-x}
\\
\sqrt{2} A(B^0 \rightarrow \pi^0 K^0)
&\equiv Y =& 2A_{3/2} - A_{1/2} + B_{1/2},
\label{isospin-formulas-y}
\\
A(B^+ \rightarrow \pi^+ K^0)
&\equiv Z =& A_{3/2} + A_{1/2} + B_{1/2},
\label{isospin-formulas-z}
\\
\sqrt{2} A(B^+ \rightarrow \pi^0 K^+)
&\equiv W =& 2 A_{3/2} - A_{1/2} - B_{1/2},
\label{isospin-formulas-w}
\end{eqnarray}
where  $A$ and $B$ represent $\Delta I = 1$ and $\Delta I = 0$ components,
respectively, and subscripts denote the isospin of final states.
These quantities satisfy the isospin relation,
\begin{eqnarray}
 X+Y=Z+W.
\label{eq:isospin_rel}
\end{eqnarray}
Decay amplitudes can be written in terms of topological contributions
\begin{eqnarray}
A(B^0 \rightarrow \pi^- K^+) &\leftrightarrow& -P - P^C_{EW} - T,
\label{diagram-rel-x}
\\
A(B^0 \rightarrow \pi^0 K^0) &\leftrightarrow&
\tfrac{1}{\sqrt{2}} P -\tfrac{1}{\sqrt{2}} P_{EW} -\tfrac{1}{\sqrt{2}} C,
\\
A(B^+ \rightarrow \pi^+ K^0) &\leftrightarrow& P + A,
\\
A(B^+ \rightarrow \pi^0 K^+) &\leftrightarrow&
- \tfrac{1}{\sqrt{2}} P
- \tfrac{1}{\sqrt{2}} P_{EW}
- \tfrac{1}{\sqrt{2}} P^C_{EW}
- \tfrac{1}{\sqrt{2}} T
- \tfrac{1}{\sqrt{2}} C
- \tfrac{1}{\sqrt{2}} A,
\label{diagram-rel-w}
\end{eqnarray}
where $T,P,P_{EW},C,P^C_{EW},P_C,A$ are ($T$) tree, ($P$) penguin, ($P_{EW}$)
electroweak (EW) penguin, ($C$) color suppressed tree,
($P^C_{EW}$) color-suppressed EW penguin,  and  ($A$) annihilation amplitude,
with strong phases $\delta'_T,\delta'_P,\delta'_{EW},\delta'_C$, $etc$., respectively.
$P$ represents the combination $P=P'-\frac{1}{3}P_{EW}^C$.

Now we introduce isospin phase
$\alpha_{3/2}'$,$\alpha_{1/2}'$ and $\beta_{1/2}'$
to each isospin component $(A_{3/2},A_{1/2},B_{1/2})$,
we get:
\begin{eqnarray}
A_{3/2}
&=& \frac{1}{3}(-T -P_{EW} - C - P^C_{EW}  ) e^{i\alpha_{3/2}'},
\label{isospinphase-a3}
\\
A_{1/2}
&=& \frac{1}{6} (-T + 2 P_{EW} + 2C - P^C_{EW} + 3A) e^{i\alpha_{1/2}'},
\label{isospinphase-a1}
\\
B_{1/2}
 &=& \frac{1}{2}(2P + T + P^C_{EW} + A ) e^{i\beta_{1/2}' },
\label{isospinphase-b1}
\end{eqnarray}
where we ignored $A$ (weak annihilation contribution) because it is estimated
to be much smaller than others within the SM evaluations of pQCD and QCD factorization.

\subsection{Branching Ratios, Direct and Mixing-Induced  CP Asymmetries}

In this subsection we express branching ratios, direct and indirect CP
asymmetries.
We remark that there exists a (naive) conventional hierarchy within the SM
among the topological diagrammatic contributions:
\begin{equation}
1 > r_{_T} \sim r_{_{EW}} > r_{_C} \sim r_{_{EW}}^C > r_{_A} ~,
\label{convhierarchy}
\end{equation}
where
$$
r_{_T} \equiv |T|/|P|,~~
r_{_{EW}} \equiv |P_{EW}|/|P|,~~
r_{_C} \equiv |C|/|P|,~~
r_{_C}^{EW} \equiv |P_C^{EW}|/|P|,~~
r_{_A} \equiv |A|/|P|.
$$
For instance, in the pQCD approach, those ratios are roughly estimated as
\cite{Mishima:2004um,Keum:2000wi}
\begin{eqnarray}
&& r_{_T} \approx 0.21,~ r_{_{EW}} \approx 0.14,~ r_{_C} \approx 0.02,
~ r_{_{EW}}^C \approx 0.01, ~ r_{_A} \approx 0.005 ~.
\label{convhierarchy2}
\end{eqnarray}
It is also known that within the SM under flavor SU(3) symmetry,
the relation $\delta^T \approx \delta^{EW}$ holds to a good approximation
\cite{Neubert:1998jq}, which can be deduced from the fact that the topology of the
color-allowed tree diagram is similar to that of the EW penguin diagram.
Here we {\it neglect} the tiny quantities $r_{_A}$ and $r_{_{EW}}^C$.
However, because recent studies on two-body hadronic $B$ decays show
that the color-suppressed tree contribution could be enhanced to a large
amount through certain mechanisms
\cite{Cheng:2004ru,Kim:2004hx,Keum:2003js}, we {\it keep} $r_{_C}$, in
order to take that possibility into account.  This treatment {\it
differs} from that in Refs. \cite{Mishima:2004um,Yoshikawa:2003hb},
where all the linear terms for $r_{_A}$ and $r_{_{EW}}^C$ as well
as $r_{_C}$ were simply neglected.
We note that
the physical strong phases, which appear in the branching ratio
and CP asymmetries after taking out overall phases, are defined to be
$$
\delta_T \equiv \delta_T' - \delta_P',~~
\delta_{EW} \equiv \delta'_{EW}-\delta'_P,~~
\delta_C \equiv  \delta'_C -\delta'_P,~~
\delta^C_{EW} \equiv \delta'^C_{EW} -\delta'_P,
$$
and
$$
\alphaone \equiv \alpha_{1/2}' - \beta_{1/2}'~~~ {\rm and}~~~
\alphathree \equiv \alpha_{3/2}' - \beta_{1/2}'.
$$

\paragraph{Branching ratios}

The CP-averaged branching ratios are given by,
\begin{eqnarray}
\bar {\cal B}^{+-} &\equiv& \bar {\cal B}(B_d \to K^{\pm} \pi^{\mp})
\nonumber\\
&\propto&
 |P|^2 |\lambdat|^2
\left\{
1
+ 2 r_{_T}  \cos \phi_3 C_{r_{_T}}^X
- 2 r_{_{EW}} C_{r_{_{EW}}}^X
\right.
\nonumber\\&&
\left.
 + r_{_T}^2 |\Lambda_X|^2 + r_{_{EW}}^2 |\Sigma_X|^2
 - 2 r_{_T} r_{_{EW}} \cos(\phi_3) C_{r_{T}r_{_{EW}}}^X
 + 2 r_{_C} C_{r_C}^X
\right\},
\label{Bpm}
\\
2 \bar {\cal B}^{+0} &\equiv& 2 \bar {\cal B}(B^{\pm} \to K^{\pm} \pi^0)
\nonumber\\
&\propto&
 |P|^2 |\lambdat|^2
\left\{
1
+ 2 r_{_T}  \cos \phi_3 C_{r_{_T}}^W
- 2 r_{_{EW}} C_{r_{_{EW}}}^W
\right.
\nonumber\\&&
\left.
 + r_{_T}^2 |\Lambda_W|^2 + r_{_{EW}}^2 |\Sigma_W|^2
 - 2 r_{_T} r_{_{EW}} \cos(\phi_3) C_{r_{T}r_{_{EW}}}^W
 + 2 r_{_C} C_{r_C}^W
\right\},
\label{Bp0}
\\
\bar {\cal B}^{0+}
&\equiv& \bar {\cal B}(B^{\pm} \to K \pi^{\pm})
\nonumber\\
&\propto&
 |P|^2 |\lambdat|^2
\left\{
1
- 2 r_{_T}  \cos \phi_3 C_{r_{_T}}^Z
+ 2 r_{_{EW}} C_{r_{_{EW}}}^Z
\right.
\nonumber\\&&
\left.
 + r_{_T}^2 |\Lambda_Z|^2 + r_{_{EW}}^2 |\Sigma_Z|^2
 - 2 r_{_T} r_{_{EW}} \cos(\phi_3) C_{r_{_T}r_{_{EW}}}^Z
 - 2 r_{_C} C_{r_C}^Z
\right\},
\label{B0p}
\\
2 \bar {\cal B}^{00} &\equiv& 2 \bar {\cal B}(B_d \to K \pi^0)
\nonumber\\
&\propto&
 |P|^2 |\lambdat|^2
\left\{
1
- 2 r_{_T}  \cos \phi_3 C_{r_{_T}}^Y
+ 2 r_{_{EW}} C_{r_{_{EW}}}^Y
\right.
\nonumber\\&&
\left.
 + r_{_T}^2 |\Lambda_Y|^2 + r_{_{EW}}^2 |\Sigma_Y|^2
 - 2 r_{_T} r_{_{EW}} \cos(\phi_3) C_{r_{T}r_{_{EW}}}^Y
 - 2 r_{_C} C_{r_C}^Y
\right\}.
\label{B00}
\end{eqnarray}
Here for simplicity we have defined parameters listed below:
\newcommand{\betaphaseprime}{e^{-i\beta_{1/2}'}}
\begin{eqnarray}
\begin{array}{lcllcl}
C_{r_{_T}}^X &\equiv& \Re \left[
\strongT \Lambda_X 
\right],
&
C_{r_{_{EW}}}^X &\equiv& \Re \left[
 \strongPEW \Sigma_X 
\right],
\\
C_{r_{_C}}^X &\equiv& \Re \left[
i \strongC \Sigma_X 
\right],
\nonumber\\
S_{r_{_T}}^X &\equiv& \Re \left[i \strongT \Lambda_X 
\right],
&
S_{r_{_C}}^X &\equiv& \Re \left[i \strongC \Sigma_X 
\right],
\\
C_{r_{_T}r_{_{EW}}}^X &\equiv&
\Re \left[  e^{i(\delta_T-\delta_{EW})} \Lambda_X \Sigma_X^* \right]
,&
S_{r_{_T}r_{_{EW}}}^X &\equiv&
\Re \left[i e^{i(\delta_T-\delta_{EW})} \Lambda_X \Sigma_X^* \right],
\end{array}
\label{eq:coeffs}
\end{eqnarray}
where
\begin{eqnarray}
\Lambda_X = e^{-i\beta_{1/2}'}
     \left(
        -\frac{1}{2}  e^{i\beta_{1/2}'}
        - \frac{1}{3}e^{i\alpha_{3/2}'}
        -\frac{1}{6}e^{i\alpha_{1/2}'}
     \right)
,&&
\Sigma_X = e^{-i\beta_{1/2}'}
     \left(
        -\frac{1}{3}e^{i\alpha_{3/2}'}
        +\frac{1}{3}e^{i\alpha_{1/2}'}
     \right),
\nonumber\\
\Lambda_Y = e^{-i\beta_{1/2}'}
     \left(
        +\frac{1}{2}  e^{i\beta_{1/2}'}
        -\frac{2}{3}e^{i\alpha_{3/2}'}
        +\frac{1}{6}e^{i\alpha_{1/2}'}
     \right)
,&&
\Sigma_Y = e^{-i\beta_{1/2}'}
     \left(
        -\frac{2}{3}e^{i\alpha_{3/2}'}
        -\frac{1}{3}e^{i\alpha_{1/2}'}
     \right),
\nonumber\\
\Lambda_Z = e^{-i\beta_{1/2}'}
     \left(
        +\frac{1}{2}  e^{i\beta_{1/2}'}
        -\frac{1}{3}e^{i\alpha_{3/2}'}
        -\frac{1}{6}e^{i\alpha_{1/2}'}
     \right)
,&&
\Sigma_Z = \Sigma_X,
\nonumber\\
 \Lambda_W = e^{-i\beta_{1/2}'}
     \left(
        -\frac{1}{2}  e^{i\beta_{1/2}'}
        -\frac{2}{3}e^{i\alpha_{3/2}'}
        +\frac{1}{6}e^{i\alpha_{1/2}'}
     \right)
,&&
 \Sigma_W = \Sigma_Y.
\nonumber
\end{eqnarray}
The CKM elements are
$\lambdat \equiv V_{tb}^* V_{ts}$, $\lambdau \equiv V_{ub}^* V_{us}$
and $\phi_3 ~ (\equiv \gamma)$ is the angle of the unitarity triangle.

\paragraph{Direct CP asymmetries}

The direct CP asymmetries are given by
\begin{eqnarray}
{\cal A}_{CP}^{+-} &\equiv&
 \frac{{\cal B}(\bar B^0 \to K^- \pi^+) -{\cal B}(B^0 \to K^+ \pi^-)}
  {{\cal B}(\bar B^0 \to K^- \pi^+) +{\cal B}(B^0 \to K^+ \pi^-)}
 = \frac{|\bar{X}|^2 - |X|^2}{|\bar{X}|^2 + |X|^2}
\label{ACPpm} \nonumber\\
&\simeq&
- 2 r_{_T} \sin \phi_3     S_{r_{_T}}^X
+ 2 r_{_T}^2 \sin(2\phi_3) S_{r_{_T}}^X C_{r_{_T}}^X
\nonumber\\&&
+ 2 r_{_T} r_{_{EW}} \sin \phi_3 S_{r_{_T} r_{_{EW}}}^X
- 4 r_{_T} r_{_{EW}} \sin \phi_3 S_{r_{_T}}^X C_{r_{_{EW}}}^X
- 2 r_{_C} \sin \phi_3 S_{r_{_C}}^X,
\\
{\cal A}_{CP}^{+0} &\equiv&
 \frac{{\cal B}(B^- \to K^- \pi^0) -{\cal B}(B^+ \to K^+ \pi^0)}
  {{\cal B}(B^- \to K^- \pi^0) +{\cal B}(B^+ \to K^+ \pi^0)}
= \frac{|\bar{W}|^2 - |W|^2}{|\bar{W}|^2 + |W|^2}
\label{ACPp0}  \nonumber\\
&\simeq&
- 2 r_{_T} \sin \phi_3 S_{r_{_T}}^W
+ 2 r_{_T}^2 \sin(2\phi_3) S_{r_{_T}}^W C_{r_{_T}}^W
\nonumber\\&&
+ 2 r_{_T} r_{_{EW}} \sin \phi_3 S_{r_{_T} r_{_{EW}}}^W
- 4 r_{_T} r_{_{EW}} \sin \phi_3 S_{r_{_T}}^Y C_{r_{_{EW}}}^W
- 2 r_{_C} \sin \phi_3 S_{r_{_C}}^W,
\\
{\cal A}_{CP}^{0+} &\equiv&
 \frac{{\cal B}(B^- \to \bar K^0 \pi^-) -{\cal B}(B^+ \to K^0 \pi^+)}
  {{\cal B}(B^- \to \bar K^0 \pi^-) +{\cal B}(B^+ \to K^0 \pi^+)}
= \frac{|\bar{Z}|^2 - |Z|^2}{|\bar{Z}|^2 + |Z|^2}
\label{ACP0p}  \nonumber\\
&\simeq&
+ 2 r_{_T} \sin \phi_3 S_{r_{_T}}^Z
+ 2 r_{_T}^2 \sin(2\phi_3) S_{r_{_T}}^Z C_{r_{_T}}^Z
\nonumber\\&&
+ 2 r_{_T} r_{_{EW}} \sin \phi_3 S_{r_{_T} r_{_{EW}}}^Z
- 4 r_{_T} r_{_{EW}} \sin \phi_3 S_{r_{_T}}^Z C_{r_{_{EW}}}^Z
+ 2 r_{_C} \sin \phi_3 S_{r_{_C}}^Z,
\label{AACP0P}
\\
{\cal A}_{CP}^{00} &\equiv&
 \frac{{\cal B}(\bar B^0 \to \bar K^0 \pi^0) -{\cal B}(B^0 \to K^0 \pi^0)}
  {{\cal B}(\bar B^0 \to \bar K^0 \pi^0) +{\cal B}(B^0 \to K^0 \pi^0)}
= \frac{|\bar{Y}|^2 - |Y|^2}{|\bar{Y}|^2 + |Y|^2}
\label{ACP00}
\nonumber\\
&\simeq&
+ 2 r_{_T} \sin \phi_3 S_{r_{_T}}^Y
+ 2 r_{_T}^2 \sin(2\phi_3) S_{r_{_T}}^Y C_{r_{_T}}^Y
\nonumber\\&&
+ 2 r_{_T} r_{_{EW}} \sin \phi_3 S_{r_{_T} r_{_{EW}}}^Y
- 4 r_{_T} r_{_{EW}} \sin \phi_3 S_{r_{_T}}^Y C_{r_{_{EW}}}^Y
+ 2 r_{_C} \sin \phi_3 S_{r_{_C}}^Y.
\end{eqnarray}

Notice that considering the conventional hierarchy given in (\ref{convhierarchy})
and (\ref{convhierarchy2}), the direct CP asymmetries ${\cal A}_{CP}^{+0}$ (\ref{ACPp0})
and ${\cal A}_{CP}^{+-}$ (\ref{ACPpm}) are expected to be almost the same including their
{\it signs}, because the dominant contribution to them is identical.
However, the current experimental data show that ${\cal A}_{CP}^{+0}$ and
${\cal A}_{CP}^{+-}$ are quite different from each other
and even have mutually opposite signs, as shown in Table~\ref{table:1}.

\paragraph{Mixing-induced  CP asymmetry}

The time-dependent CP asymmetry for $B^0 \to K_{_S} \pi^0$ is defined as
\begin{eqnarray}
{\cal A}_{K_{_S} \pi^0} (t)
 &\equiv& \frac{\Gamma(\bar B^0 (t) \to K_{_S} \pi^0) -\Gamma(B^0 (t) \to K_{_S} \pi^0)}
  {\Gamma(\bar B^0 (t) \to K_{_S} \pi^0) +\Gamma(B^0 (t) \to K_{_S} \pi^0)}  \nonumber \\
 &\equiv& S_{K_{_S} \pi^0} \sin(\Delta m_d ~t) +C_{K_{_S} \pi^0} \cos(\Delta m_d ~t),
\end{eqnarray}
where $\Gamma$ denotes the relevant decay rate and $\Delta m_d$ is the mass
difference between the two $B^0$ mass eigenstates.
$S_{K_{_S} \pi^0}$ and $C_{K_{_S} \pi^0}$ are CP violating parameters.
In the case that the tree contributions are neglected for $B^0 \to K_{_S} \pi^0$,
the mixing-induced CP violating parameter $S_{K_{_S} \pi^0}$ is equal to $\sin(2\phi_1)$
[$\phi_1 ~ (\equiv \beta)$ is the angle of the unitarity triangle].
The expression for $S_{K_{_S} \pi^0}$ (up to $r$ order) is given by
\begin{eqnarray}
S_{K_{_S} \pi^0}
\label{Skpi}
&=& \Im\left(- e^{-2i\phi_1}
  \frac{\bar{Y}/\lambdat^*}{Y/\lambdat}\right)
\nonumber\\
&\simeq& \sin 2\phi_1
- 2r_{_T} \sin \phi_3
 \Re \left(e^{-2i\phi_1} \strongT \Lambda_Y 
\right)
- 2 r_{_C} \sin \phi_3
 \Re \left(e^{-2i\phi_1} \strongC \Sigma_Y 
\right).
\end{eqnarray}
The measured value of $S_{K_{_S} \pi^0}$ (Table~\ref{table:1}) is different from the
well-established value of $\sin(2\phi_1) =0.725 \pm 0.037$ measured through
$B \to J/\psi K^{(*)}$ \cite{HFAG}.
It may indicate that the subleading terms including
$r_{_C}$ and $r_{_{EW}}$ in Eq. (\ref{Skpi})
play an important role. In Ref. \cite{Kim:2005jp}, the authors showed that as $S_{K_{_S} \pi^0}$
varies, the allowed region for $r_{_C}$ varies very sensitively,
but that for $r_{_W}$  not,
as can be seen from Eq. (\ref{Skpi}).

\section{The $B \to K \pi$ puzzle and its physical implications}

\subsection{Summary of Present Experimental Results}

We first summarize the present status of the experimental results on $B \to K\pi$
modes in Table I, which includes the BRs, the direct CP asymmetries
$({\cal A}_{CP})$, and the mixing-induced CP asymmetry $(S_{K_s \pi^0})$.
We see that the averages of the current experimental values for the BRs include
only a few percent errors.
Furthermore, the direct CP asymmetry in $B^0 \to K^{\pm} \pi^{\mp}$ has been recently
observed by the BaBar and Belle collaborations whose values
are in good agreement with each other (Table I): the world average value is
\begin{equation}
{\cal A}_{CP}^{+-} = -0.109 \pm 0.020 ~.
\end{equation}
The direct CP asymmetry data for the other $B \to K\pi$ modes involve large
uncertainties.

\begin{table}
\caption{Experimental data on the CP-averaged branching ratios ($\bar {\cal B}$
in units of $10^{-6}$), the direct CP asymmetries (${\cal A}_{CP}$), and
the mixing-induced CP asymmetry ($S_{K_s \pi^0}$) for $B \to K\pi$ modes.
The $S_{K_s \pi^0}$ is equal to $\sin(2\phi_1)$ in the case that tree
amplitudes are neglected for $B^0 \to K_s \pi^0$
\cite{Bornheim:2003bv,Chao:2003ue,Aubert:2002jb,Aubert:2004dn,Aubert:2004km,Aubert:2004kn,
Chen:2000hv,Aubert:2004qm,Chao:2004jy,Chao:2004mn,Abe:2004xp}.}
\smallskip
\begin{tabular}{|c|c|c|c|c|}
\hline
  & CLEO & Belle & BaBar & Average  \\
\hline
$\bar {\cal B}(B^{\pm} \to K^0 \pi^{\pm})$ & $18.8^{+3.7 +2.1}_{-3.3 -1.8}$
 & $22.0 \pm 1.9 \pm 1.1$ & $26.0 \pm 1.3 \pm 1.0$ & $24.1 \pm 1.3$  \\
$\bar {\cal B}(B^{\pm} \to K^{\pm} \pi^0)$ & $12.9^{+2.4 +1.2}_{-2.2 -1.1}$
 & $12.0 \pm 1.3^{+1.3}_{-0.9}$ & $12.0 \pm 0.7 \pm 0.6$ & $12.1 \pm 0.8$  \\
$\bar {\cal B}(B^0 \to K^{\pm} \pi^{\mp})$ & $18.0^{+2.3 +1.2}_{-2.1 -0.9}$
 & $18.5 \pm 1.0 \pm 0.7$ & $17.9 \pm 0.9 \pm 0.7$ & $18.2 \pm 0.8$  \\
$\bar {\cal B}(B^0 \to K^0 \pi^0)$ & $12.8^{+4.0 +1.7}_{-3.3 -1.4}$
 & $11.7 \pm 2.3^{+1.2}_{-1.3}$ & $11.4 \pm 0.9 \pm 0.6$ & $11.5 \pm 1.0$  \\
\hline
${\cal A}_{CP}^{0+}$ & $0.18 \pm 0.24$ & $0.05 \pm 0.05 \pm 0.01$
 & $-0.087 \pm 0.046 \pm 0.010$ & $-0.020 \pm 0.034$   \\
${\cal A}_{CP}^{+0}$ & $-0.29 \pm 0.23$ & $0.04 \pm 0.05 \pm 0.02$
 & $0.06 \pm 0.06 \pm 0.01$ & $0.04 \pm 0.04$   \\
${\cal A}_{CP}^{+-}$ & $-0.04 \pm 0.16$ & $-0.101 \pm 0.025 \pm 0.005$
 & $-0.133 \pm 0.030 \pm 0.009$ & $-0.109 \pm 0.020$   \\
${\cal A}_{CP}^{00}$ & $-$ & $-0.12 \pm 0.20 \pm 0.07$
 & $-0.06 \pm 0.18 \pm 0.06$ & $-0.09 \pm 0.14$   \\
\hline
$S_{K_s \pi^0}$ & $-$ & $0.30 \pm 0.59 \pm 0.11$
 & $0.35^{+0.30}_{-0.33} \pm 0.04$ & $0.34 \pm 0.29$  \\
\hline
\end{tabular}
\label{table:1}
\end{table}

We comment on the values of the ratios between the BRs for the $B \to K\pi $ modes,
$R_1$, $R_c$, and $R_n$, which are obtained from the experimental results given in Table I:
\begin{eqnarray}
&& R_1 =  \frac{\tau^+ \bar{\cal B}^{+-}}{\tau^0 \bar{\cal B}^{0+}}=0.82 \pm 0.06 ~,
\label{R1:data} \\
&& R_c = \frac{2\bar{\cal B}^{+0}}{\bar{\cal B}^{0+}}= 1.00 \pm 0.09 ~,
\label{Rc:data} \\
&& R_n = \frac{\bar{\cal B}^{+-}}{2\bar{\cal B}^{00}}= 0.79 \pm 0.08 ~.
\label{Rn:data}
\end{eqnarray}
It has been  claimed that within the SM, $R_c - R_n \approx 0$
\cite{Buras:2004th,Buras:2003dj}.
{}From their definitions, it is indeed clear that $R_c \approx R_n$,
if the $r^2$-order terms including $r_{_{EW}}$ or $r_{_C}$ are negligible.
In other words, any difference between $R_c$ and $R_n$ would arise
from the contributions
from the subdominant $r^2$-order terms including $r_{_{EW}}$ or $r_{_C}$.
The above experimental data show the pattern $R_c > R_n$ \cite{Buras:2004th,Buras:2003dj},
which would imply an enhancement of the electroweak penguin and/or the color-suppressed tree
contributions.

We note that in this analysis we do not consider $B \to \pi \pi$ modes simultaneously
with $B \to K\pi$ modes, though
they can be connected to each other by using flavor SU(3) symmetry.
The reason is that  we do not want that our analysis to be spoiled
by the unknown effects of the flavor SU(3) breaking relations
after the inclusion of FSI effects.

We remind that assuming the conventional hierarchy as in Eqs. (\ref{convhierarchy})
and (\ref{convhierarchy2}), ${\cal A}_{CP}^{+0}$ is expected to be almost
the same as ${\cal A}_{CP}^{+-}$: in particular, they would have the {\it same} sign.
However, the data show that ${\cal A}_{CP}^{+0}$ differs by 3.4$\sigma$ from
${\cal A}_{CP}^{+-}$.
This is a very interesting observation with the new measurements of
${\cal A}_{CP}^{+-}$ by BaBar and Belle, even though the measurements of
${\cal A}_{CP}^{+0}$ still include sizable errors.
One may need to explain on the theoretical basis how this feature can arise.

\subsection{Global $\chi^2$ Analysis and Theoretical Implications}

\newcommand{\chimin}{\chi_\mathrm{min}^2}
\newcommand{\dof}{\mathrm{d.o.f}}

Based on the current experimental data shown in Table I, we critically
investigate their implications to the underlying theory on the $B \to K\pi$
processes.
There are nine observables available for the $B \to K\pi$ modes, as shown
in Table~\ref{table:1}.
However, there are  ten theoretical parameters ($|P|$, $r_{_T}$, $r_{_{EW}}$, $r_{_C}$,
$\delta^T$, $\delta^{EW}$, $\delta^C$, $\alpha_{1/2}$, $\alpha_{3/2}$ and
$\phi_3(\equiv \gamma$ of the unitary triangle))
relevant to the above nine
observables, neglecting the very small terms of the annihilation contribution\footnote{
After the annihilation term $r_{_A}$
is neglected, the observable ${\cal A}_{CP}^{0+}$  still remains non-zero
due to non-zero values of the isospin phases.
In our approach, all nine observables  still remain relevant.} $r_{_A}$,
and the color suppressed electroweak penguin  contribution $r_{_{EW}}^C$.
We have fixed $\sin2\phi_1(\equiv \beta) = 0.726$
to its central value of the experimental measurements.
Therefore, we have to fix at least  one of the theoretical input parameters
by assuming a model-calculated value or  the previously measured central value
in order to do $\chi^2$ analysis.

\subsubsection{\bf Global $\chi^2$ Analysis within Isospin-phased Topological Approach}

%
\begin{table}[ht]
\caption{
$\chi^2$ fit for {\bf cases(a-d)} and their combinations, including all
 sdSPs and ldSPs. Fixed parameters in each column  appear in
 parentheses.}
\label{g-fit}
\small
\begin{tabular}{|c||c|c|c|c|c|c|}
\hline
 & {\bf (a)} & {\bf (b)} & {\bf (a+b)} & {\bf (c)} & {\bf (d)} & {\bf (c+d)} \\
\hline
$\chimin/\dof$& $0.0/0$         & $0.0/0$        &$0.0/1$&
                $0.1/0$         & $2.02/0$       &$2.04/1$ \\
\hline
$p$           & $22.7_{\pm2.3}$ & $23.56_{\pm0.78}$  &$22.9_{\pm0.80}$&
                $21.9_{\pm1.8}$ & $24.0_{\pm1.3}$    &$23.8_{\pm1.3}$
\\
$\alphaone$   & $-0.03_{\pm0.41}$ &$-0.10_{\pm0.25}$ & $-0.05_{\pm0.27}$ &
                $ 0.52_{\pm0.39}$ &$ 0.25_{\pm0.71}$ & $ 0.35_{\pm0.25}$
\\
$\alphathree$ & $ 0.23_{\pm0.71}$ & $0.0_{\pm0.35}$    & $0.18_{\pm0.36}$ &
                $-1.04_{\pm0.68}$ & $3.048_{\pm0.095}$ & $3.04_{\pm0.087}$
\\
$\delta_T$    & $0.31_{\pm0.46}$   & $0.32_{\pm0.13}$ & $0.28_{\pm0.11}$ &
                $0.128_{\pm0.046}$ & $0.20_{\pm0.10}$ & $0.19_{\pm0.11}$
\\
$\delta_{EW}$ & $1.68_{\pm0.69}$ & $2.00_{\pm0.34}$ & $1.73_{\pm0.29}$ &
                ($=\delta_T$)    & $0.55_{\pm2.0}$  & ($=\delta_T$)
\\
$\delta_{C}$  & $-3.01_{\pm0.43}$ & $-2.90_{\pm0.29}$ & $-3.01_{\pm0.35}$ &
                $-3.92_{\pm0.65}$ & ($=\delta_P$)     & ($=\delta_P$)
\\
$r_{_T}$      & $0.19_{\pm0.24}$  & ($0.21$)         & ($0.21$) &
                $0.58_{\pm0.18}$  & $0.33_{\pm0.14}$ & $0.31_{\pm0.11}$
\\
$r_{_{EW}}$   & $0.35_{\pm0.13}$  & $0.322_{\pm0.080}$ & $0.347_{\pm0.087}$&
                $0.01_{\pm0.33}$  & $0.16_{\pm0.20}$   & $0.142_{\pm0.040}$
\\
$r_{_C}$      & $0.28_{\pm0.24}$  & $0.30_{\pm0.19}$   & $0.29_{\pm0.21}$
              & $0.54_{\pm0.28}$  & $0.00_{\pm0.15}$   & $0.00_{\pm0.15}$
\\
$\phi_3$($^o$)&($60.0$)         & $50.9_{\pm5.5}$  & ($60.0$)
              & $73.3_{\pm8.4}$ & $64.0_{\pm10.0}$ & $64.8_{\pm10.4}$
\\
\hline
\end{tabular}
\end{table}
We do the $\chi^2$ analysis  to investigate sources of physics
beyond the SM step-by-step as follows: for 9 observables (4 branching ratios,
4 direct CP asymmetries and one indirect CP asymmetry) with 9 input parameters
by fixing {\bf case(a)}: $\phi_3=60^0$ (but
$p \propto \kappa |P|^2|\lambda_t|^2$,
$\alphaone$, $\alphathree$, $\delta_{_{T,EW,C}}$
and $r_{_{T,EW,C}}$ are free parameters.), or {\bf case(b)}:
$r_{_T}=0.21$ (the pQCD central value) or {\bf case(c)}:
$\delta_T=\delta_{EW}$ or {\bf case(d)}: $\delta_C=0$ ($i.e.$
$\delta'_C=\delta'_P$).  We further exercise by fixing {\bf case(e)}:
the values of all sdSPs zero (typical isospin analysis), where we have
three fewer parameters, or {\bf case(f)}: the values of all ldSPs zero
(usual topological quark diagrammatic approach), in which we have two
fewer parameters; and combinations of {\bf cases(a-f)}.

In Table \ref{g-fit}, we show the results of $\chi^2$-fitting of 9
observables by parameters including both ldSPs and sdSPs.
For the minimum of $\chi^2$, in {\bf cases(a, b, c, a+b)} $\chimin$
value are almost zero, whereas in {\bf cases(d)} $\chimin$ becomes larger than 2,
which means that $\delta_C=\delta_P$ for sdSPs is not a good assumption.
{}From now on, we do not consider the {\bf case(d)} any more, except the
combination with {\bf case(c)} which assumes $\delta_C=\delta_P$ as well as
$\delta_T=\delta_{EW}$.
We searched the whole range of parameter spaces for global $\chimin$,
and we obtained the large ldSPs:  $\alphaone$ in the range $-0.9 < \alphaone < 0.5$ (in the
unit of radian) and $\alphaone = 0$ is allowed within the range of
$1\sigma$ in {\bf cases(a, b, d, a+b)}; and
for $\alphathree$, in {\bf cases(a, b, a+b)} $\alphathree=0$ is allowed
within the $1\sigma$ level; whereas in {\bf case(c)} $\alphathree=0$ is not
allowed in $1\sigma$.
Central values of $\delta_T$ are distributed in the range $ 0.7^{o} <
\delta_T < 18^{o} $, and
central values of $\delta_{EW}$ are large and about $96^{o} < \delta_{EW}
< 115^{o}$ in {\bf cases(a, b, a+b)}; however, the assumption $\delta_T=\delta_{EW}$
is quite valid with larger $r_{_T}$, much larger than typical pQCD estimates.
Central values of $\delta_C$ are very large and almost $180^{o}$ in all
cases with the exception of {\bf cases(d, c+d)}
in which $\delta_C=\delta_P$ is assumed.
For $r_{_T}$, in {\bf cases(a, b, a+b)} central values are in good agreement with
its pQCD value $r_{_T}=0.21$.
As for $r_{_{EW}}$ and $r_{_C}$, in general large values are preferred  for better
$\chi^2$ fitting;
central values of $r_{_{EW}}$ are around 0.35, much larger than its typical
pQCD value $r_{_{EW}} = 0.14$, in {\bf cases(a, b, a+b)}.
For $r_{_C}$, {\bf cases(a, b, c, a+b)} exclude the pQCD value
$r_{_C}=0.018$ by more than $1\sigma$.

%
%
\begin{table}[htb]
\caption{ CP asymmetries estimated from  best-fit parameters for each
case in TABLE \ref{g-fit}.  }
\label{CP-fit}
\footnotesize
\begin{tabular}{|l||c|c|c|c|c|c|}
\hline
CP asym. (Exp.) &
 {\bf (a)} & {\bf (b)}& {\bf (a+b)}
 & {\bf (c)} & {\bf (d)} & {\bf (c+d)} \\
\hline
${\cal A}_{CP}^{+-} $ ($-0.109_{\pm0.020}$)
 & $-0.109_{\pm0.28}$ & $-0.109_{\pm0.067}$ & $-0.109_{\pm0.080}$
 & $-0.107_{\pm0.33}$ & $-0.106_{\pm0.100}$ & $-0.106_{\pm0.038}$
\\
${\cal A}_{CP}^{+0} $  ($0.04_{\pm0.04}$)
 & $0.04_{\pm0.34}$ & $0.04_{\pm0.17}$ & $0.04_{\pm0.21}$
 & $0.04_{\pm0.48}$ & $0.026_{\pm0.085}$ & $0.025_{\pm0.043}$
\\
${\cal A}_{CP}^{0+} $  ($-0.020_{\pm0.034}$)
 & $-0.020_{\pm0.11}$ & $-0.020_{\pm0.055}$ & $-0.020_{\pm0.070}$
 & $-0.021_{\pm0.30}$ & $-0.014_{\pm0.122}$ & $-0.013_{\pm0.053}$
\\
${\cal A}_{CP}^{00} $  ($-0.09_{\pm0.14}$)
 & $-0.09_{\pm0.28}$ & $-0.09_{\pm0.16}$ & $-0.08_{\pm0.20}$
 & $-0.09_{\pm0.57}$ & $-0.18_{\pm0.21}$ & $-0.19_{\pm0.11}$
\\
${\cal S}_{K_S\pi^0}$  ($0.34_{\pm0.29}$)
 & $0.34_{\pm0.38}$ & $0.34_{\pm0.26}$ & $0.33_{\pm0.32}$
 & $0.43_{\pm0.30}$ & $0.09_{\pm0.28}$ & $0.12_{\pm0.05}$
\\
\hline
\end{tabular}
\end{table}
In Table \ref{CP-fit},
we show the estimated CP asymmeries from the best fit parameters for each
case in Table \ref{g-fit}, to be compared with the experimental average values of
Table \ref{table:1}.  We do not show the estimated branching ratios because
there are almost no differences between the experimental values and the theoretically
estimated values from the best fit parameters.

A few comments are in order here:
(i) Contrary to the previous finding \cite{Mishima:2004um,Kim:2005jp}
within the topological approach,
{\bf case(c)} confirms $\delta_T=\delta_{EW}$ of Ref. \cite{Neubert:1998jq}.
It can be understood by the fact that the topology of the
tree diagram is quite similar to that of the EW penguin diagram.
(ii) {\bf Case(c+d)} is very interesting since it gives the pQCD estimated values on
the parameters, $r_{_T},~r_{_{EW}},~r_{_C}$ and $\phi_3$ (see Eq. (\ref{convhierarchy2})).
That is, if we include FSI, we can fit the parameters within the SM
without invoking any unknown new physics effects at all.
(iii) The ldSPs from FSI can be quite large, and therefore, we cannot ignore
the final state re-scattering effects even in $B$ meson decays.
(iv) Indeed, it has been long advocated that charming-penguin contributions can
increase significantly the $B \to K \pi$ rates and yield better agreement
with experiment \cite{charming-penguin}. Strong phase $\delta_c$ from charming-penguin
has been also found substantially large, $\delta_c \sim 20^\circ$ \cite{charm-phase}.
(v) Long distance strong phases has been modelled  \cite{Cheng:2004ru}
as re-scattering  of some intermediate two-body states with one particle exchange
in the $t$ channel and the absorptive
part of the re-scattering amplitude via optical theorem. Large long distance
contributions are found as $\delta_T - \delta_C \sim 90^\circ$
with FSI, too.
(vi) It has been found \cite{Donoghue} that by using general features of
soft strong interactions soft scattering does not decrease for large $m_B$,
and inelastic processes are expected to be leading sources of strong
phases\footnote{
>From the viewpoint of Regge theory, it is pointed
out \cite{Donoghue} that even in the heavy quark limit there are
non-vanishing inelastic non-perturbative FSI effect remaining on each
individual channels.
However, in \cite{Beneke:2000ry} authors showed the inelastic FSI
effects will be cancelled by each other when all individual channels are
summed up.
Because in $B$ hadronic decays the $b$-quark mass is finite, therefore,
such a cancellation might be incomplete and the
non-perturbative inelastic effects could be sizable \cite{Cheng:2005wf}.  }.
Please
note that in this paper we only include the elastic FSI contributions
through isospin phases.

\par 

\subsubsection{\bf Null Hypothesis Test by Statistical $p-$value}

Here we make a hypothetical test by using a statistical $p-$value\footnote{
In the sense of statistics, the $p-$value is the probability that the
observed value  happens to be observed under the ``null
hypothesis".
When the $p-$value is smaller than a threshold (usually 5\% is used), it can
be said that the null hypothesis is disfavored and the ``alternative
hypothesis" is favored.},
which is defined by the integral $\int_{\chimin}^{\infty} \! g(t;d) dt$,
where $\chimin$
is the minimum $\chi^2$ value obtained and $g(t;d)$ the probability density
function for $\chi^2$ distribution with given $d$ degree of freedom.
One hypothesis which we want to test is the
assumption that the parameters $r_{T}$, $r_{_{EW}}$ and $r_{C}$ take their pQCD
estimated values, $e.g.$ $r_{_T} \sim
0.21$, $r_{_{EW}} \sim 0.14$ and $r_{_C} \sim 0.018$. We
call this the ``amplitude assumption".
The other hypothesis is the relation among sdSPs: $\delta_T =\delta_{EW}$
and $\delta_C=\delta_P$,
which we call the ``strong phase assumption".

\begin{table}[tbp]
\caption{The $\chi^2$-result with fixed $r$'s ($i.e.$ $r_{_T} = 0.21$, $r_{_{EW}} = 0.14$
and $r_{_C} = 0.018$).
In ''free" case, $p$, $\alphaone$, $\alphathree$, $\delta_{T,EW,C}$ and $\phi_3$ are all
free parameters. In each cases 2 ldSPs, 4 branching ratio, 4 direct and
 1 indirect CP
 assymmetries at the $\chi^2$-minimum are also shown.}
\label{hypo}
\begin{center}
\small
\begin{tabular}{|l||c|c|c|c|c|}
\hline
Strong Phases & \multicolumn{2}{|c|}{both ldSPs and sdSPs} &
 \multicolumn{2}{|c|}{only sdSPs} & only ldSPs \\
\hline
 & {\bf (a)} + free $\phi_3$  & {\bf (c+d)}
 & {\bf (f)} & {\bf (f+c)} & {\bf (e)} \\
\hline
\hline
$\chimin/\dof$ & $2.3/2$ & $4.3/3$ & $8.4/4$ & $12.8/5$ & $16.2/5$ \\
$p-$value(\%)    & $32$    & $23$    & $7.8$   & $2.5$    & $0.6$
\\
\hline
\end{tabular}
\end{center}
\end{table}
In Table \ref{hypo}, we summarize the $\chimin$ and its $p-$value
with amplitude and strong phase assumptions for each
case $-$ case with only ldSPs, case with only sdSPs and case with both ldSPs and
sdSPs.
In the case with both ldSPs and sdSPs, $p-$value is about 32\% with
non-constrained sdSPs (free), and one with constrained sdSPs ({\bf
case(c+d)}) is still around 23\%.
%
In the case with only sdSPs ({\bf cases(f, f+c)}), $p-$value in {\bf (f)} is
7.8\%, while in {\bf (f+c)} the probability is reduced to 2.5\% and this
case is almost ruled out.
%
In the case with only ldSPs ({\bf case(e)}), the $p-$value is only
0.6\% and this case must be completely ruled out.
Therefore, this test confirms our previous claim that the generalized
``isospin-phased topological approach" can
fit the $B \to K \pi$ decays well with the SM values of the parameters, which
are estimated from pQCD,
without invoking  any unknown new physics effects.
We again conclude that we cannot ignore
the final state re-scattering effects and ldSPs even in $B$ meson decays.

\subsubsection{\bf Dependence of  $S_{K_S\pi}$ and ${\cal A}_{CP}^{+0}$
on $r_{_{EW}}$ and $r_{_C}$}

\begin{table}[tbp]
\caption{$\chi^2$ result for {\bf (A)} varying $S_{K_S\pi^0} = 0.2,\,0.4,\,0.6,\,0.7$,  and,
{\bf (B)} varying ${A_{CP}^{+0}} = +0.04,\,+0.02,\,-0.02,\,-0.04$.
Error is assumed to be 20\%. Concerning the input parameters, we choose {\bf case(a)}
($i.e.$ ldSPs
 and sdSPs are included, $\phi_3=60^o$ is fixed, and $\sin2\beta = 0.726$ is used).}
\label{tbl-dep}
\begin{center}
\small
{\bfseries (A)}\\
\begin{tabular}{lllll} \hline
\multicolumn{1}{|l|}{$S_{K_S\pi^0}$} &
 \multicolumn{1}{l|}{0.20$\pm$0.04} &
 \multicolumn{1}{l|}{0.40$\pm$0.08} &
 \multicolumn{1}{l|}{0.60$\pm$0.12} &
 \multicolumn{1}{l|}{0.70$\pm$0.14} \\
 \hline
\hline
\multicolumn{1}{|l|}{$\chimin$} &
 \multicolumn{1}{r|}{0.0} &
 \multicolumn{1}{r|}{0.0} &
 \multicolumn{1}{r|}{0.0} &
 \multicolumn{1}{r|}{0.0} 
\\
 \hline
\multicolumn{1}{|l|}{$r_{_{EW}}$} &
 \multicolumn{1}{l|}{0.34$\pm$0.16} &
 \multicolumn{1}{l|}{0.36$\pm$0.12} &
 \multicolumn{1}{l|}{0.36$\pm$0.10} &
 \multicolumn{1}{l|}{0.36$\pm$0.10} 
\\
 \hline
\multicolumn{1}{|l|}{$r_{_C}$} &
 \multicolumn{1}{l|}{0.39$\pm$0.09} &
 \multicolumn{1}{l|}{0.23$\pm$0.09} &
 \multicolumn{1}{l|}{0.08$\pm$0.07} &
\multicolumn{1}{l|}{0.06$\pm$0.11}
\\
 \hline
\end{tabular}
  \end{center}
\begin{center}
\small
{\bfseries (B)}\\
\begin{tabular}{lllll} \hline
\multicolumn{1}{|l|}{${\cal A}_{CP}^{+0}$} &
\multicolumn{1}{r|}{$+0.04\pm0.008$} &
\multicolumn{1}{r|}{$+0.02\pm0.004$} &
\multicolumn{1}{r|}{$-0.02\pm0.004$} &
\multicolumn{1}{r|}{$-0.04\pm0.008$} \\
 \hline
\hline
\multicolumn{1}{|l|}{$\chimin$} &
 \multicolumn{1}{r|}{0.0} &
 \multicolumn{1}{r|}{0.0} &
 \multicolumn{1}{r|}{0.0} &
 \multicolumn{1}{r|}{0.0} \\
 \hline
\multicolumn{1}{|l|}{$r_{_{EW}}$} &
 \multicolumn{1}{l|}{$0.35\pm0.13$} &
 \multicolumn{1}{l|}{$0.30\pm0.16$} &
 \multicolumn{1}{l|}{$0.38\pm0.16$} &
 \multicolumn{1}{l|}{$0.38\pm0.10$} \\
 \hline
\multicolumn{1}{|l|}{$r_{_C}$} &
 \multicolumn{1}{l|}{$0.28\pm0.25$} &
 \multicolumn{1}{l|}{$0.28\pm0.24$} &
 \multicolumn{1}{l|}{$0.15\pm0.09$} &
 \multicolumn{1}{l|}{$0.26\pm0.26$}
\\ \hline
\end{tabular}
\end{center}
\end{table} %

Now we would like to make  a few comments on the sensitivity
of the observable $S_{K_S\pi^0}$ to the parameter $r_{_C}$.
The experimental value of  $S_{K_S\pi^0}$ is
much smaller than that of $S_{J/\psi K_S}$.
As implied by (\ref{Skpi}), the theoretical prediction of
$S_{K_S\pi^0}$ can be very sensitive to the parameter $r_{_C}$. For
illustration, in part {\bf (A)} of Table \ref{tbl-dep}, we
vary $S_{K_S\pi^0}$ around the present experimental value, specifically,
$S_{K_S\pi^0}$ is assumed to be $(0.20\pm0.04)$ , $(0.40\pm0.08)$ ,
$(0.60\pm 0.12)$ and $(0.70\pm 0.14)$, respectively.
Here just for the illustrative purpose,
we set $20\%$ errors in each case. (Also, to be consistent, we set
$20\%$ errors to all the data whose current errors are larger than
$20\%$, such as $A_{CP}^{ij}$.)
It can be seen that as $S_{K_S\pi^0}$ varies, the allowed region for
$r_{_C}$ varies in a clearly correlated manner.
Just for comparison, in  {\bf (B)} of Table \ref{tbl-dep}, we also present the
case when the value of $A_{CP}^{+0}$ varies.
We recall that the averaged experimental value of $A_{CP}^{+0}$ is {\it positive},
in contrast to the value of $A_{CP}^{+-}$.
Again for the illustrative purpose, $A_{CP}^{+0}$ is assumed to be
$(\pm0.04\pm0.008)$,$(\pm0.02\pm0.004)$, respectively.
(To be consistent, we also set $20\%$ errors to all the data whose
current errors are larger than $20\%$, such as $S_{K_S\pi^0}$ and
$A_{CP}^{+0}$.)
In contrast to case {\bf (A)},
the dependence of $r_{_{EW}}$ and $r_{C}$  on $A_{CP}^{+0}$ is very obscure.
%

\section{Conclusions}

All CP-averaged branching ratios of $B \to K\pi$ modes have been recently
measured within a few percent errors by the BaBar and Belle collaborations.
The observations of the direct CP asymmetry in $B^0 \to K^{\pm}
\pi^{\mp}$ have also been recently achieved at the 5.7$\sigma$ level by BaBar
and Belle.
Certain experimental data
($e.g.$, the branching ratios (BRs)) for $B \to K\pi$ are currently more
precise than the theoretical model predictions based on QCD
factorization or perturbative QCD.
Thus, with the
currently available precision data, it is very important to investigate
these modes as generally and critically as possible.
The importance of the soft final state re-scattering effects, especially, in
a model-independent topological quark diagrammatic analysis of $B \to D
\pi$, $B \to K \pi$ and $B \to \pi \pi$ decays, has also
been recognized recently.
It is certainly conceivable that final state interactions can modify
the predictions based on
the short distance diagrammatic analysis.
In this work, we studied the decay processes $B \to K\pi$ by using
generalized decay amplitudes within the so-called ``isospin-phased
topological approach", $i.e.$ with both the strong phases
of the final state re-scattering and the strong phases of
the topological quark diagrammatic origin.
We are mainly interested in investigating whether the conventional
SM predictions are consistent with the current data even after we
include final state interaction effects.
In order to find the most likely values of the parameters
in a statically reliable way, we used the $\chi^2$ analysis.

Our result shows that:
(i) Contrary to the previous finding within the topological approach,
we confirmed $\delta_T=\delta_{EW}$ of Ref. \cite{Neubert:1998jq}, which
can be understood by the fact that the topology of the
tree diagram is quite similar to that of the EW penguin diagram.
(ii) If we include final state interaction effects, we can fit
theoretical input parameters within the SM
without invoking any unknown new physics effects at all.
(iii) The final state re-scattering  phases can be quite large,
and therefore, we cannot ignore
the final state re-scattering effects even in $B$ meson decays.
(iv) We also found that there are strong correlations between the parameter
$r_{_C}$ and the time dependent indirect CP observable $S_{K_s \pi^0}$:
If  $S_{K_s \pi^0} \approx S_{J/\psi K_s}$, then $r_{_C} < 0.1$.
However, the present experimental value, $S_{K_s \pi^0} \ll S_{J/\psi K_s}$,
directly implies very large $r_{_C}$ like $0.3 \sim 0.4$,
if final state re-scattering phases are not considered.
\\


\centerline{\bf ACKNOWLEDGEMENTS}
\noindent
We would like to thank G. Cvetic for careful reading of the manuscript
and his valuable comments.
The work of CSK was supported
by the Korea Research Foundation Grant funded by the Korean Government
(MOEHRD) No. KRF-2005-070-C00030, and by JSPS.
He also wishes to thank Physics Division of National Center for
Theoretical Sciences(NCTS) at Hsinchu, ROC for hospitality during his
visit.
The work of DC, CSC and HH has been supported in part by a grant from
National Science Foundation of ROC and in part by NCTS.
DC has been recently deceased of stomach cancer.
\\


\end{document}